\documentclass{modified_ijmpe}

\usepackage{color}

\begin{document}

\markboth{Bally, Avez, Bender, and Heenen}
         {Symmetry restoration for odd-mass nuclei}

\catchline{}{}{}{}{}

\title{SYMMETRY RESTORATION FOR ODD-MASS NUCLEI WITH A SKYRME ENERGY DENSITY FUNCTIONAL
       \footnote{Based on talk presented at 18th Nuclear Physics Workshop "Maria and Pierre Curie", 2011, Kazimierz, Poland.}}

\author{B. Bally, B. Avez, and M. Bender}
\address{Universit{\'e} Bordeaux 1, CNRS/IN2P3,
             Centre d'Etudes Nucl{\'e}aires de Bordeaux Gradignan,
             CENBG, Chemin du Solarium, BP120,
             F-33175 Gradignan, France}

\author{P.-H. Heenen}
\address{PNTPM, CP229,
             Universit{\'e} Libre de Bruxelles,
             B-1050 Bruxelles,
             Belgium}

\maketitle


\begin{abstract}
In these proceedings, we report first results for particle-number and 
angular-momentum projection of self-consistently blocked triaxial 
one-quasiparticle HFB states for the description of odd-$A$ nuclei 
in the context of regularized multi-reference energy density functionals, using 
the entire model space of occupied single-particle states. The 
SIII parameterization of the Skyrme energy functional and a volume-type
pairing interaction are used.
\end{abstract}
%
%
\section{Introduction}

Methods based on the self-consistent mean-field approach
provide a set of fully microscopic theoretical tools that can
be applied to all bound atomic nuclei in a systematic manner
irrespective of their mass, $N/Z$ ratio, and deformation.\cite{RMP}
Pure mean-field methods, however, have several limitations. The first 
one is due to the determination of a wave function in an intrinsic
frame of reference. Although the symmetry-breaking mean-field
approach is a very efficient and transparent way to grasp
the effect of correlations associated with collective modes in
the limit of strong correlations,\cite{Rin80a,BlaRip} the absence of good
quantum numbers and the corresponding selection rules compromises 
the calculation of transition moments.
For example, broken rotational symmetry mixes states with different 
eigenvalues of $\hat{J}^2$, i.e.\ the members of a rotational band, 
and independent quasiparticle states of BCS type used to describe 
pairing correlations are spread over several particle numbers.
The second limitation concerns nuclei for which a mean-field
description through a single configuration breaks down because several 
configurations with different shell structure are close in energy.
%
%
To overcome these limitations requires what is often called 
"beyond-mean-field methods", i.e.\ symmetry restoration and
configuration mixing within the Generator Coordinate 
Method (GCM).\cite{Rin80a,BlaRip,BadHonnef,LesHouches}

There are many more applications of mean-field and ``beyond'' 
methods to even-even nuclei than there are for odd-$A$ nuclei 
(not to mention odd-odd ones). 
The most important reason is that the last nucleon in the odd-$A$ system
breaks some of the symmetries that make such methods particularly
efficient for even-even systems.
%
%
\section{Odd-mass nuclei in self-consistent mean-field models}
Odd-mass nuclei are described as one-quasiparticle states build 
on a HFB vacuum.\cite{Rin80a} The most transparent way to represent 
such a state is in the canonical basis $\{ a^\dagger_i \}$
\begin{equation}
| q_b \rangle
= a^\dagger_b
  \prod_{i \neq b > 0}    
  \big(   \bar{U}_{i i}
        + \bar{V}_{i \bar\imath} a^\dagger_i a^\dagger_{\bar\imath}
  \big) \, | - \rangle 
\end{equation}
where $i$ and $\bar\imath$ are the indices of the paired conjugate
states, where we use the usual convention $i > 0$ and $\bar\imath < 0$,
and $b$ is the index of the blocked single-particle state. The 
single-particle states and Bogoliubov matrices $U$ and $V$ are determined 
self-consistently. The corresponding HFB equation is obtained from variation of  
%
%
the energy with the obligatory constraints ensuring that $| q_b \rangle$
remains a quasiparticle vacuum and on proton and neutron 
number.\cite{Rin80a,BlaRip} Optional 
constraints are on the two components of the quadrupole tensor in the 
intrinsic major axis system of the triaxial shape and on the angular 
momentum component $I_x$ along one major axis. As done in our earlier 
configuration mixing calculations, we add the Lipkin-Nogami prescription 
to the HFB equations to enforce pairing correlations in all mean-field 
states.\cite{Gal94a,Rig99a} The resulting HFB equations for the blocked 
state are solved using the "two-basis method", where in an iterative 
procedure the HFB Hamiltonian is diagonalized in the single-particle 
basis of eigenstates of the HF Hamiltonian in a coordinate-space 
representation.\cite{Gal94a}



%
%
\section{Going ``beyond the mean field''}
Restoring the symmetries decomposes a given SCMF state into states 
with good quantum numbers. Eigenstates of the particle-number 
operator $\hat{N}$ with eigenvalue $N_0$ are obtained applying 
the particle-number projection operator\cite{BlaRip}
\begin{equation}
\hat{P}_{N_0}
= \frac{1}{2 \pi}
  \int_{0}^{2 \pi} \! d \phi_N \;
  e^{i \phi_N (\hat{N}-N_0)}
\end{equation}
to the SCMF states for protons and neutrons.
Eigenstates of the total angular-momentum operator in the 
laboratory frame $\hat{J}^2$ and its $z$ component 
$\hat{J}_z$ with eigenvalues \mbox{$\hbar^2 J (J+1)$} and 
$\hbar M$, respectively, are obtained applying 
the operator\cite{BlaRip}
\begin{equation}
\label{eq:PJ}
\hat{P}^J_{MK}
= \frac{2J+1}{16 \pi^2}
  \! \int_{0}^{2\pi} \! \! d\alpha
  \! \int_0^\pi \! \! d\beta \; \sin(\beta)
  \!  \int_0^{4 \pi} \! \! d\gamma \;
  \mathcal{D}^{J \, \ast}_{MK} \, (\alpha,\beta,\gamma) \,
  \hat{R} (\alpha,\beta,\gamma)
\, ,
\end{equation}
where \mbox{$\hat{R} = e^{-i\alpha \hat{J}_z}
e^{-i\beta \hat{J}_y} e^{-i\gamma \hat{J}_z}$} is the rotation
operator and $\mathcal{D}^{J}_{MK}$ a Wigner function. 
Both depend on the Euler angles $\alpha$, $\beta$, and $\gamma$. 
$\hat{P}^J_{MK}$
picks the component with angular-momentum projection $K$ along the
intrinsic $z$ axis from a mean-field state. The projected state is then 
obtained by summing over all $K$ components
\begin{equation}
\label{eq:Kmix}
| J M {\nu} q_b \rangle
= \sum_{K=-J}^{+J} f_{\nu}^{J} (K,q_b) \, \hat{P}^J_{MK} \, | q_b \rangle
= \sum_{K=-J}^{+J} f_{\nu}^{J} (K,q_b) \, | J M K q_b \rangle
\end{equation}
with weights $f_{\nu}^{J}(K,q_b)$ determined from a variational principle that leads 
to the so-called Hill-Wheeler-Griffin (HWG) equation\cite{Rin80a,BlaRip}
\begin{equation}
\label{eq:HWG}
\sum_{K'=-J}^{+J}
\Big( \langle J M K q_b | \hat{H} | J M K' q_b \rangle
      - E_\nu \, \langle J M K q_b | J M K' q_b \rangle \Big) \; 
    f^{J}_{\nu} (K',q_b)
= 0
\, ,
\end{equation}
%
%
%
which for sake of simple notation we write for a Hamilton operator.
In practice, we use an effective interaction that is provided 
by a multi-reference (MR) energy density functional (EDF). 
It is the generalization of the single-reference 
(SR) EDF employed in SCMF methods to calculate the non-diagonal 
kernels entering the symmetry-restored energy and the HWG 
equation~(\ref{eq:HWG}). It is usually postulated in a form that preserves 
analogies with the case of a Hamilton operator.\cite{Bon90a,Rod02b,Rob07a} 
However, the MR EDF has to be regularized to remove spurious contributions 
to the energy that manifest themselves through divergences and/or finite 
steps when plotting the symmetry-restored energy as a function of a 
collective coordinate.\cite{I,II} We have implemented such regularization\cite{IVa,IV}
into our codes for general configuration mixing.\cite{Ben08a}
We use the parameterization SIII\cite{Bei75a} of the Skyrme EDF together 
with a contact pairing interaction of volume type and a strength of
300~MeV~fm$^3$ together with cutoffs at 5~MeV above and below the Fermi 
energy.\cite{Rig99a} The Coulomb exchange term is neglected as done 
in our earlier regularized MR EDF calculations.\cite{II,IVa}
The non-diagonal norm kernels entering the symmetry-restored
HWG equation~(\ref{eq:HWG}) are calculated directly with their sign
in a technique based on a Pfaffian.

%
%
\section{An illustrative example: $^{49}$Cr}

We have chosen $^{49}$Cr for a first exploratory study as it exhibits
several low-lying collective rotational bands that can be easily 
associated with particular blocked single-particle levels. A detailed 
study of the adjacent even-even $^{48}$Cr that allows for the analysis of 
the change brought by the additional nucleon is also underway.
%
%
\subsection{Projected energy surfaces}

 The first step of our calculation was to determine which quasiparticle gives the lowest energy at the Single-Reference level. It is
 a quasiparticle with
 parity minus and with the mean value of the projection along the deformation axis of the angular momentum close to $\frac52$ :
 ${\langle J_3\rangle}^{\pi} \approx 2.5^{-}$, which agrees with the experimental finding for the ground state that has J$^\pi$ = $\frac52^-$. We 
 then tried to follow this quasiparticle in the first sextant of the $\beta-\gamma$ plane. As we have neither good  $J$ nor good $J_3$, however, it is
 not always
 clear which quasiparticle should be blocked. This ambiguity presents an additional motivation to go to the Multi-Reference level, that is to restore the 
 angular momentum. 
 As demonstrated for example by Schunck \emph{et al.},\cite{NSc10a} because
 of the time-odd components in the EDF the sextants are not equivalent, and a full study would require to explore at least three out of the six.
 Test calculations in the other sextants indicate, however, that for this blocked quasiparticle the differences after K mixing remain tiny, and do not
 have any importance for the present discussion. This might not be the case for other blocked quasiparticles, and certainly there will be significant
 differences between the sextants when cranking the quasiparticle states to higher spin.

\begin{figure}[bth!]
 \centerline{\epsfig{file=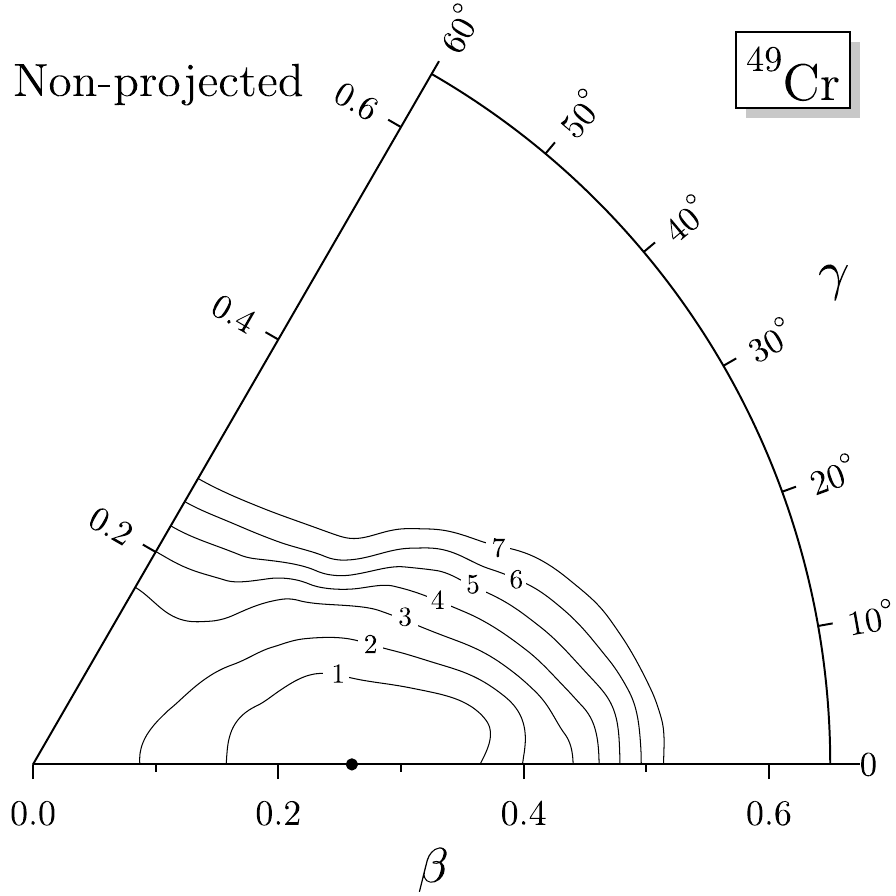 ,width=06.3cm}
             \epsfig{file=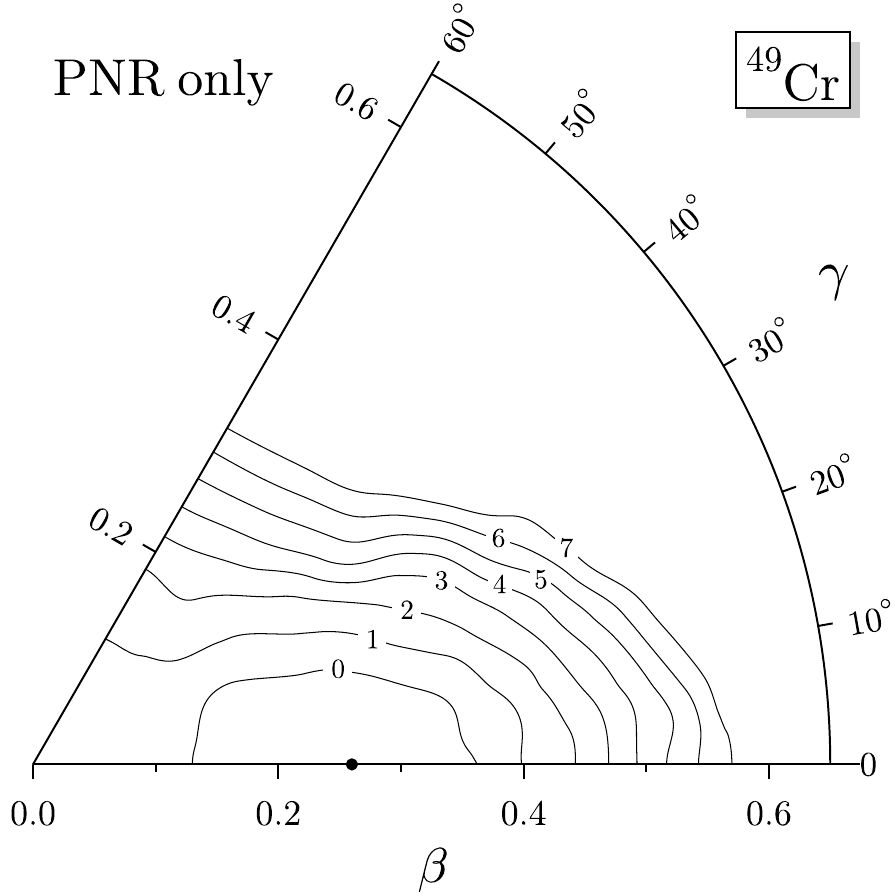   ,width=06.3cm}}
 \centerline{\epsfig{file=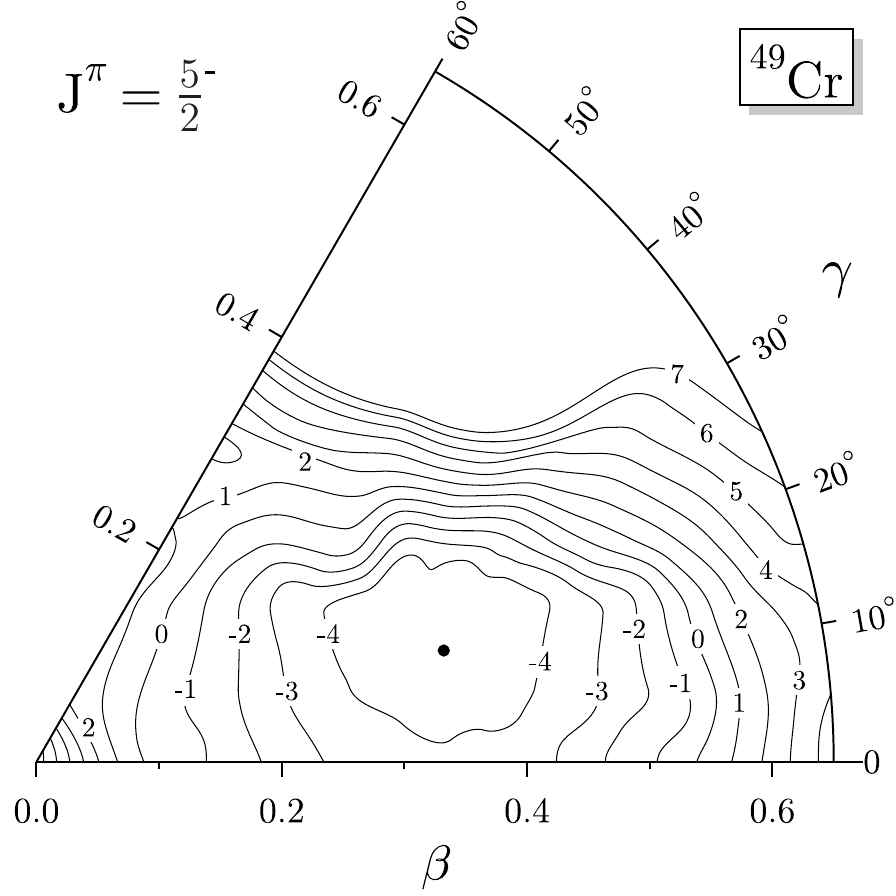  ,width=06.3cm}}
\caption{\label{fig:PJNZ}
Top left: Non-projected energy surface, in the first sextant of the $\beta-\gamma$ plane, of the lowest one-quasiparticle state with ${\langle J_3\rangle}^{\pi} \approx 2.5^{-}$ in $^{49}$Cr.
Top right: Particle-number restored energy surface constructed from the same one-quasiparticle states.
Bottom: Particle-number and angular-momentum restored energy surface for the lowest J$^\pi$ = $\frac52^-$, based again on the same one-quasiparticle states.
Energies are normalized to the minimum of the non-projected surface. The deformation parameter $\beta = \sqrt{\frac {5}{16}}\frac {4\pi}{3R^{2}A}\langle Q_2 \rangle$ with $R \equiv 1.2 A^{1/3}$ fm and $\gamma$ are those of the (non-projected) intrinsic states.
}
\end{figure}

 In the top left of Fig.~\ref{fig:PJNZ} is plotted the non-projected energy surface up to 7 MeV. The energy is normalized to the minimum 
 we found and which turns out to be axial ($\beta = 0.26$, $\gamma = 0.0^{\circ}$). In the top right of Fig.~\ref{fig:PJNZ} we present the 
 particle number restored (PNR) energy surface, 
 where the projected energies are plotted at the deformation of the SCMF states they have been obtained from.\footnote{The same representation is also
 adopted for the angular-momentum restored surfaces.} One observes that the PNR does not 
 change the global shape of the surface. In particular, the minimum remains axial. The PNR gives an almost constant overall energy gain of about 1 MeV 
 for the entire surface.
 When projecting on both particle number and angular momentum, we
 get for J = $\frac52$, the J which gives the absolute minimum after projection, the surface plotted in the bottom of Fig.~\ref{fig:PJNZ}. 
 And as one can see, the topography of the surface has
 dramatically changed: the minimum has moved into triaxality ($\beta = 0.33$, $\gamma = 15.3^{\circ}$), and the surface is now centered around this minimum,  which is about 5 MeV lower than the non-projected one.
 The angular-momentum projected surface is thus more rigid against oblate deformation than the non-projected one.

\begin{figure}[bth!]
 \centerline{\epsfig{file=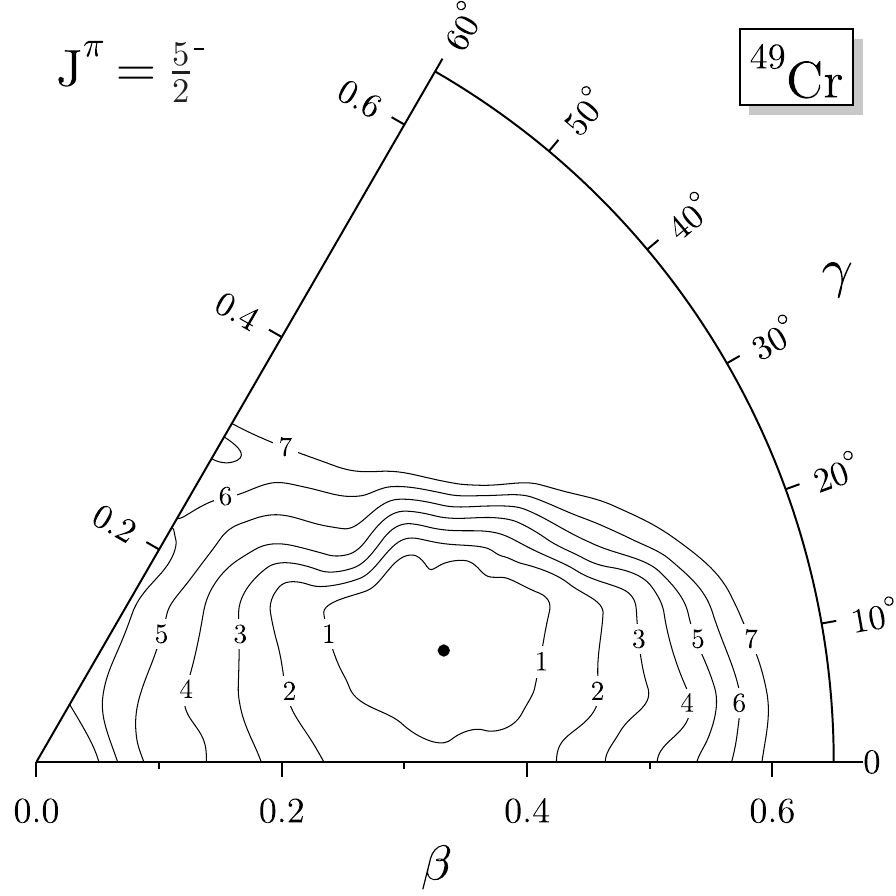 ,width=06.3cm}
             \epsfig{file=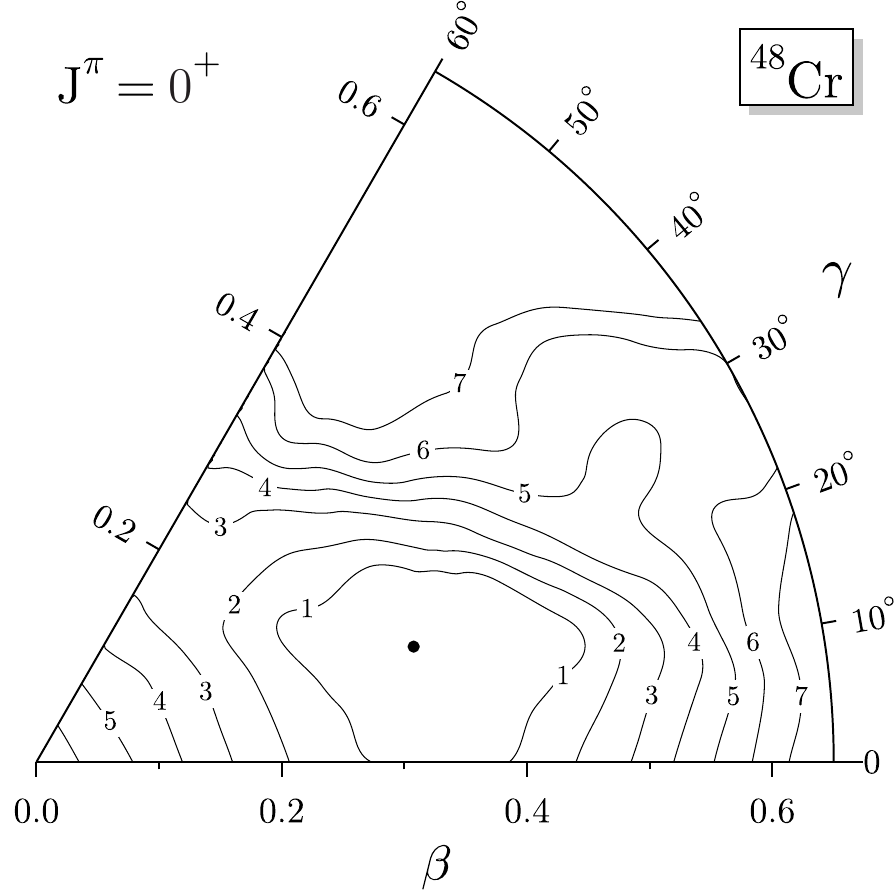 ,width=06.3cm}}
\caption{\label{fig:pairimpair}
Left: Energy surface, in the first sextant of the $\beta-\gamma$ plane, of the lowest J$^\pi$ = $\frac52^-$ after K mixing in $^{49}$Cr. 
Right: Same for the lowest J$^\pi$ = $0^+$ in $^{48}$Cr. Energies are normalized to the minimum of each surface, which is marked by a filled circle.
}
\end{figure}

 In Fig.~\ref{fig:pairimpair} we compare the J$^\pi$ = $\frac52^-$ energy surface of $^{49}$Cr (left), now renormalized to its minimum, with the J$^\pi$ = $0^+$
 energy surface of $^{48}$Cr (right), also normalized to its minimum. One immediatly sees that the deepest part of the surface is very similar in both 
 cases, in particular the minimum does not move much by the addition of one extra nucleon to $^{48}$Cr. By contrast, the energy
 surface
 of the J$^\pi$ = $\frac52^-$ in $^{49}$Cr is clearly more rigid than the J=$0^+$ surface in $^{48}$Cr, i.e. for the former the energy grows faster when moving away from the minimum. It will be interesting to see if these
 remarks remain true for different J coming from other blocked quasiparticles.

\begin{figure}[tbh!]
 \centerline{\epsfig{file=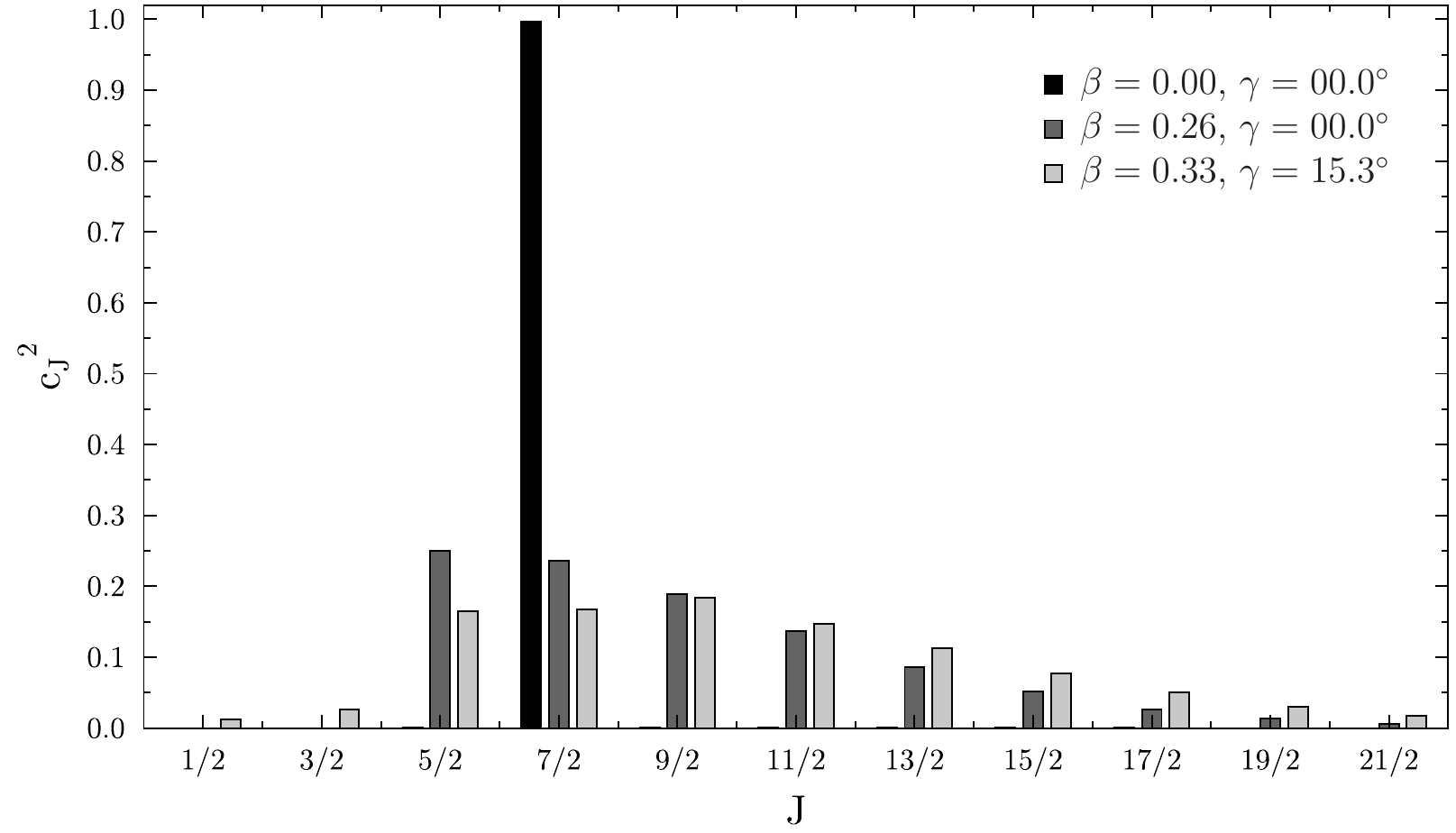 , width=10.0cm}}
 \vspace*{8pt}
 \centerline{\epsfig{file=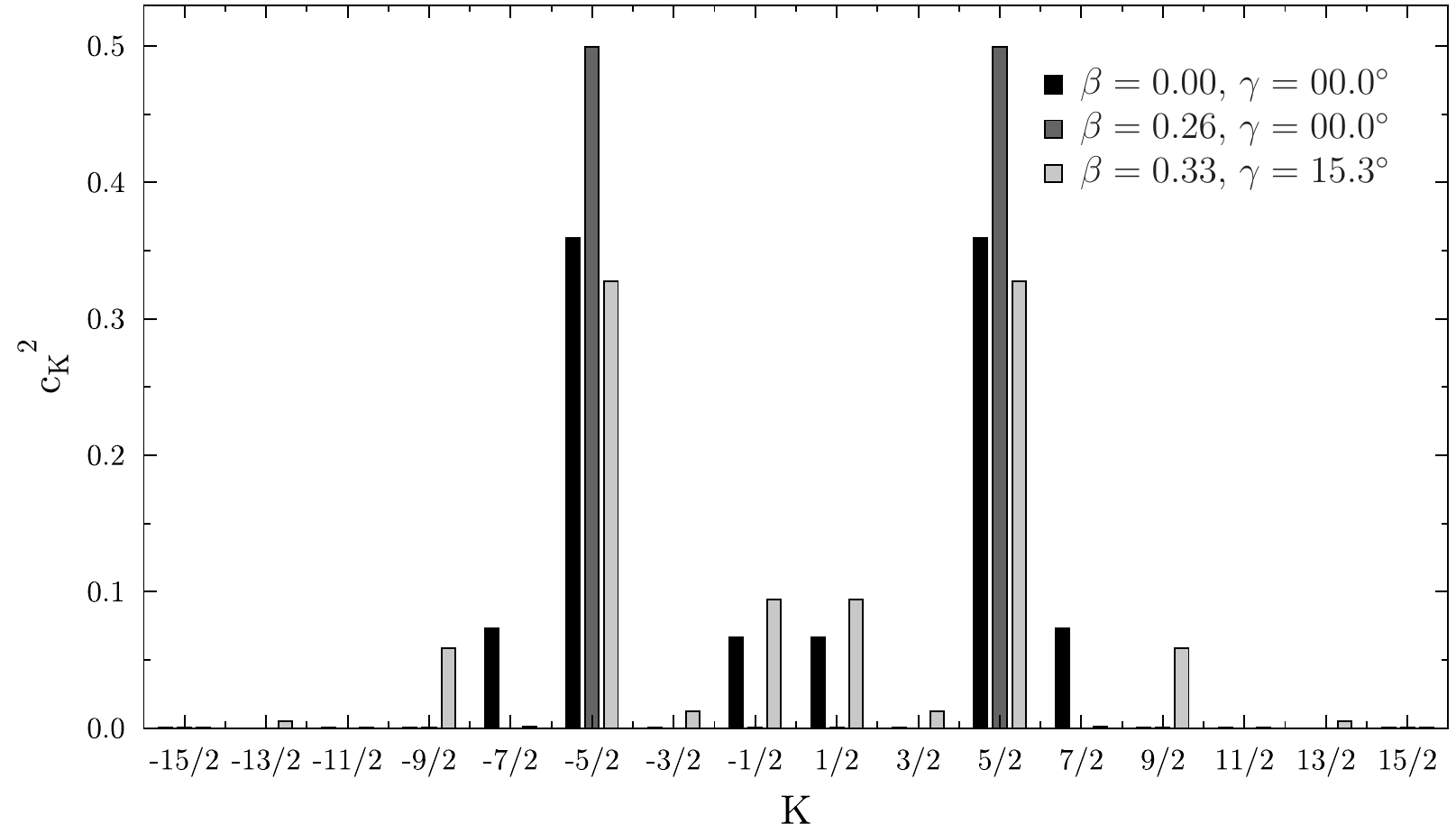 , width=10.0cm}}
 \vspace*{8pt}
 \caption{\label{fig:JK_axial}
 Top: square of the weights of the different J components summed over K, for three different deformations.
 Bottom: same but for the weights of the K components summed over J.
 }
\end{figure}

%
%
\subsection{J and K decompositions of SCMF states}

  In order to better understand how the projection of quasiparticle states works, it is illuminating to look at the decomposition of SCMF states
 on different angular momenta J and their intrinsic components K. We plot in Fig.~\ref{fig:JK_axial} the square of the weights of J and K components
 for several points in the $\beta$-$\gamma$ plane with different deformation. If one looks first at the axially constrained point that corresponds to the minimum before projection (dark
 grey bars in Fig.~\ref{fig:JK_axial}), one sees that there is indeed not just one J, but a full distribution beginning at J = $\frac52$, which is the 
 largest component in the distribution, and decreasing with increasing J. Together with the gain in energy, the fact that there are different J 
 components in the SCMF state proves  a posteriori the need for angular-momentum projection. One also notes that there is no component with 
 J = $\frac12$ or with  J = $\frac32$. 
 This can be understood when looking at the K decomposition of this state. Here, one can see that there are only two different K components, K = 
 $\frac52$ and its partner\footnote{Finding components of equal weight for $\pm$K is a consequence of our choice of having eigenstates of the $x$ signature operator:
$\hat{R_x} = e^{-i\pi\hat{J_x}}$.}
  K = $-\frac52$. As K has to be less than or equal to J, the J = $\frac12$ and J = $\frac32$ components
 vanish. It is also interesting to note that even if we don't have axial symmetry, a state constrained to axial deformation of its local density has
 in first approximation a good K. \\
 The light grey bars of Fig.~\ref{fig:JK_axial} represent the J and K decomposition of the state which gives the triaxial minimum after projection 
 on J = $\frac52$. As one can see, the distribution has changed: it is now the J = $\frac92$ component which is the largest, and the distribution
 is more evenly divided around this maximum. But more importantly, the K decomposition 
 one has now not only K = $\pm\frac52$, but a distribution of K with a staggering which favors  K = $\pm\frac12,\pm\frac52,\pm\frac92 \dots$ 
 So even if  K = $\pm\frac52$ remains the largest component in this distribution, this points out the necessity for a K mixing as presented in
  Eqns.~(\ref{eq:Kmix}) and (\ref{eq:HWG}). 

 Finally, the black bars in Fig.~\ref{fig:JK_axial} represent the J and K decomposition of a state constrained to a spherical density distribution.
 One notices that there is only a J = $\frac72$ component
 in the decomposition. This can be easily understood: imposing spherical density, the single-particle wave functions are almost eigenstates of angular
 momentum, such that the HFB state is also almost an eigenstate of J, with the eigenvalue determined by the j of the blocked particle, which in this case is
 in the  $f_{7/2}$ shell. By contrast, nothing in our calculation fixes which magnetic substate in the $f_{7/2}$ shell is blocked. As a consequence,
 we find an arbitrary combination that depends mainly on the initialization of the blocked HFB calculation.

%
%
\section{Summary and Outlook}

We have reported first results obtained from a method that describes
properties of odd-$A$ nuclei by particle-number and angular-momentum
projected configuration mixing based of self-consistently blocked
one-quasiparticle states and using the Skyrme EDF. 

We focussed here on symmetry restoration in a regularized MR-EDF framework. A symmetry-restored GCM 
calculation along the same lines is underway and will be reported 
elsewhere.\cite{Bal11x} Having a method that allows to treat 
odd-$A$ and even-even nuclei on the same footing provides a versatile 
tool for numerous studies of nuclear structure, such as the coupling 
of single-particle and shape degrees of freedom, the analysis of 
signatures for shell structure such as separation energies, $g$ factors, 
spectroscopic quadrupole moments, and their evolution with $N$ and $Z$,
or the analysis of the interplay of pairing correlations and fluctuations 
in shape degrees of freedom for the odd-even mass staggering.

%
%
\section*{Acknowledgments}
This research was supported in parts by the Agence Nationale 
de la Recherche under Grant No.~ANR 2010 BLANC 0407 "NESQ",
by the IN2P3/CNRS through the PICS No.~5994, and the PAI-P6-23 
of the Belgian Office for Scientific Policy. The computations 
were performed using HPC resources from GENCI-IDRIS (Grant 2011-050707).
BB also thanks for travel support by LEA-COPIGAL.
%
%


\begin{thebibliography}{99}

\bibitem{RMP}
M. Bender, P.-H. Heenen, P.-G. Reinhard,
\textit{Rev. Mod. Phys.} \textbf{75} (2003) 121.

\bibitem{Rin80a}
P. Ring and P. Schuck, 
\textit{The Nuclear Many-Body Problem},
(Springer, New York, Heidelberg, Berlin, 1980).

\bibitem{BlaRip}
J. P. Blaizot and G. Ripka,
\textit{Quantum theory of finite systems}
(MIT, Cambridge, 1986).

\bibitem{BadHonnef}
J. L. Egido and L. M. Robledo,
Lect. Notes Phys. \textbf{641} (Springer, Berlin, 2004), 269.

\bibitem{LesHouches}
M. Bender,
\textit{Eur. Phys. J.} \textbf{ST156} (2008) 217.

\bibitem{Gal94a}
B. Gall \emph{et al.},
\textit{Z. Phys. A} \textbf{348} (1994) 183.

\bibitem{Rig99a}
C. Rigollet, P. Bonche, H. Flocard, and P.-H. Heenen,
\textit{Phys. Rev.} C \textbf{59} (1999) 3120.

\bibitem{Bon90a}
P. Bonche \emph{et al.},
\textit{Nucl. Phys. A} \textbf{510} (1990) 466.

\bibitem{Rod02b}
R. R. Rodr{\'i}guez-Guzm{\'a}n \emph{et al.},
\textit{Nucl. Phys. 1} \textbf{709} (2002), 201.

\bibitem{Rob07a}
L. M. Robledo,
\textit{Int. J. Mod. Phys. E} \textbf{16} (2007) 337.

\bibitem{I}
D. Lacroix, T. Duguet, and M. Bender,
\textit{Phys. Rev. C} \textbf{79} (2009) 044318.

\bibitem{II}
M. Bender, T. Duguet, and D. Lacroix,
\textit{Phys. Rev. C} \textbf{79} (2009) 044319.

\bibitem{IVa}
M. Bender, T. Duguet, P.-H. Heenen, D. Lacroix,
\textit{Int. J. Mod. Phys. E} \textbf{20} (2011) 259.

\bibitem{IV}
M. Bender, B. Avez, T. Duguet, P.-H. Heenen, and D. Lacroix,
\textit{in preparation}.

\bibitem{Ben08a}
M. Bender and P.-H. Heenen,
\textit{Phys. Rev.} C \textbf{78} (2008) 024309.

\bibitem{Bei75a}
M. Beiner \emph{et al.},
\textit{Nucl. Phys.} \textbf{A238} (1975) 29.

\bibitem{Ave11x}
B. Avez and M. Bender, 
preprint arXiv:1109.2078 v1.

\bibitem{NSc10a}
  N. Schunck \emph{et al.},
\textit{Phys. Rev.} C \textbf{81} (2010) 024316.

\bibitem{Bal11x}
B. Bally, B. Avez, M. Bender, and P.-H. Heenen,
\emph{in preparation}.


\end{thebibliography}
\end{document}